\documentclass[sigconf]{acmart}

\AtBeginDocument{%
  \providecommand\BibTeX{{%
    \normalfont B\kern-0.5em{\scshape i\kern-0.25em b}\kern-0.8em\TeX}}}

\setcopyright{acmlicensed}
\copyrightyear{2025}
\acmYear{2025}
\acmDOI{XXXXXXX.XXXXXXX}

\acmConference[]{ }{}{}

\acmISBN{978-1-4503-XXXX-X/18/06}

\usepackage{soul}
\usepackage{times}
\usepackage{empheq}
\usepackage{latexsym}
\usepackage{booktabs}
\usepackage{multirow}
\usepackage{subfigure}
\usepackage{tabularx}
\usepackage{color}
\usepackage{graphicx}
\usepackage{caption}
\usepackage{amsmath}

\usepackage[ruled,vlined]{algorithm2e}
\usepackage{algorithmic}
\usepackage{float}

\SetKwInput{KwInput}{Input}
\SetKwInput{KwOutput}{Output}

\begin{document}

\title{Learning To Sample the Meta-Paths for Social Event Detection}

\author{Congbo Ma}
\email{congbo.ma@mq.edu.au}
\affiliation{%
  \institution{Macquarie University}
  \city{Sydney}
  \state{NSW}
  \country{Australia}
}

\author{Hu Wang}
\email{hu.wang@mbzuai.ac.ae}
\affiliation{%
  \institution{MBZUAI}
  \city{Abu Dhabi}
  \country{United Arab Emirates}
}

\author{Zitai Qiu}
\email{zitai.qiu@mq.edu.au}
\affiliation{%
  \institution{Macquarie University}
  \city{Sydney}
  \state{NSW}
  \country{Australia}
}

\author{Shan Xue}
\email{emma.xue@mq.edu.au}
\affiliation{%
  \institution{Macquarie University}
  \city{Sydney}
  \state{NSW}
  \country{Australia}
}

\author{Jia Wu}
\email{jia.wu@mq.edu.au}
\affiliation{%
  \institution{Macquarie University}
  \city{Sydney}
  \state{NSW}
  \country{Australia}
}

\author{Jian Yang}
\email{jian.yang@mq.edu.au}
\affiliation{%
  \institution{Macquarie University}
  \city{Sydney}
  \state{NSW}
  \country{Australia}
}

\author{Preslav Nakov}
\email{preslav.nakov@mbzuai.ac.ae}
\affiliation{%
  \institution{MBZUAI}
  \city{Abu Dhabi}
  \country{United Arab Emirates}
}

\author{Quan Z. Sheng}
\email{michael.sheng@mq.edu.au}
\affiliation{%
  \institution{Macquarie University}
  \city{Sydney}
  \state{NSW}
  \country{Australia}
}

\begin{abstract}

Social event detection involves identifying and analyzing events on social media platforms, which is crucial for various applications such as disaster management, public safety monitoring and marketing strategies. Social media data is inherently rich, as it includes not only text content, but also users, geolocation, entities, temporal information, and their relationships. This data richness can be effectively modeled using heterogeneous information networks (HINs) as it can handle multiple types of nodes and relationships, allowing for a comprehensive representation of complex interactions within social data. Meta-path-based methods use the sequences of relationships between different types of nodes in an HIN to capture the diverse and rich relationships within the social networks.
However, the performance of social event detection methods is highly sensitive to the selection of meta-paths and existing meta-path based detectors either rely on human efforts or struggle to determining the effective meta-path set for model training and evaluation. In order to automatically discover the most important meta-paths, we propose a simple, yet effective, end-to-end Learning To Sample (LTS) framework for meta-path searching. Specifically, we build graphs that contain not only user profiles, textual content, and details about entities, but also the intricate relationships among them. The prioritized meta-paths, based on their importance, are sampled from the maintained distribution and their features are constructed before feeding into the social event detector. After picking up the top-ranked meta-paths, we streamline the exponential increment of meta-path combinations into a finite set of highly influential ones. The chosen meta-paths, along with their respective weights, are then used to train our social event detection model. As an alternative to social event detector training, we further propose an extra non-parametric evaluation process in order to determine the importance of each meta-path, which can further guide the paths sampling during model training.  
\end{abstract}



\maketitle

\section{Introduction}
Social event detection, which refers to the task of automatically identifying and categorizing events within online social media platforms,  received increasing attention in the recent years \cite{hao2021streaming, Ren2022From, Cao2023Hierarchical, Ren2022Evidential, Cao2021Knowledge, Liu2020Story}. These events range from real-world happenings such as natural disasters \cite{pekar2020early}, future prediction \cite{liu2019embedding}, political elections, or sport events to more everyday occurrences such as social gatherings, parties, or cultural festivals. Awareness of event occurrence is vital for the responses on emergency management, business intelligence, public sentiment analysis, or community engagement. 

Importantly, social media data contain not only textual information (e.g., posts, comments), but also temporal information, metadata (e.g., user IDs, geolocations) and network structure (relations between users). This heterogeneous makes detecting social events very challenging. Heterogeneous Information Networks (HINs) \cite{sun2012mining} have the capability to incorporate different types of nodes and edges, representing diverse entities and relationships, and thus can integrate various types of social data into a unified network. This heterogeneity enhances the depth of information in representing social connections, providing a more comprehensive view of the social network \cite{shi2016survey}. However, most social event detection approaches \cite{Cao2021Knowledge, peng2022reinforced, Ren2022From, qiu2024heterogeneous} transfer the HINs into a homogeneous structure by aggregating all attribute embeddings and then using Graph Attention Networks \cite{Velickovic2018Graph} for neighbor aggregation. This simplification overlooks the structural and semantic information inherent in HINs, posing challenges for extracting meaningful features, which are essential for accurate event detection. For instance, if we treat user nodes and message nodes as the same type and aggregate their attribute embeddings, we lose crucial information specific to each node type. In social networks, user nodes and text nodes have distinct attributes; for example, users may have attributes such as gender, geographic location, and number of followers, while posts may have attributes such as text content, posting time, and number of likes. Aggregating these different types of attributes into a single vector results in a loss of valuable information specific to each type.
Furthermore, GNN-based models still face two key limitations: they cannot adequately account for the varying qualities of connections within graphs across different perspectives \cite{Ren2022Evidential}; and the sparsity inherent in social event data may hinder GNN-based models from effectively leveraging the available information for making accurate predictions.

In order to leverage the rich heterogeneity information in HINs and to make the model more robust to sparse data, one popular way is to extract meta-paths from HINs. Meta-path based methods, by defining meta-paths that capture diverse relationships between nodes in the graph, can help alleviate the issues caused by data sparsity and improve the model's robustness. But how to select the meta-path is still a problem. Some methods incorporate experts' experience to define meta-path \cite{Sun2013PathSelClus, Peng2019Fine, wang2019heterogeneous, hao2021streaming, peng2022reinforced}, but such methods heavily rely on human effort. To overcome this issue, some methods \cite{Seongjun2019Graph, li2021graphmse} incorporate all meta-paths, but this is redundant and may lead to low efficiency \cite{sun2009ranking}. Some meta-paths carry interferential information \cite{Liu2017Event, wei2018unsupervised}. According to Li et al. \cite{Li2023long}, certain meta-paths dominate model performance, while others contribute less. Selecting a suitable set of meta-paths is a challenging task in social event detection.

To address the aforementioned issues when modeling rich social media data and selecting suitable meta-path sets, in this paper, we propose an end-to-end meta-path \underline{\textbf{L}}earning \underline{\textbf{T}}o \underline{\textbf{S}}ample (LTS) framework. It aims to identify information-rich and important meta-paths that can most accurately detect social events through a limited search. We first model the social data based on HINs, and the proposed LTS integrates event elements including keywords, geolocation, entities, temporal information, social platform users, and the relationship between them using a semantically meaningful set of meta-path features for model training. We then propose five distinct sampling strategies aimed at identifying an appropriate set of meta-paths for subsequent evaluation and updating: 
(1) Random Sampling, (2) Multinomial Distribution Sampling, (3) Decayed $\epsilon$-greedy Sampling, (4) Multinomial $\epsilon$-greedy Sampling, and (5) Multinomial Encouraged $\epsilon$-greedy Sampling. 
Moreover, in order to accurately evaluate the importance of different meta-path and to guide the path sampling process, we propose a non-parametric weighted accumulation evaluation process to initialize, evaluate, update, and normalize the meta-path contributions based on the given metric. Directly linking the meta-path importance to the metric that ought to be optimized.
Our main contributions can be summarized as follows: 

\begin{itemize}
\item To extract the most important meta-paths, we propose a simple, yet effective, end-to-end meta-path searching framework: Learning To Sample (LTS) for social event detection. The best meta-paths can be effectively and efficiently sampled for model training, based on their importance.
\item Besides social event detector training, we propose a non-parametric meta-path searching technique to further guide path sampling. In the search process, we further propose different path sampling and weight normalization methods.
\item The proposed LTS model achieves state-of-the-art results across datasets for social event detection. We further offer extensive analysis, including different meta-path importance weight normalisation, as well as qualitative and quantitative analyses of the impact for the top-ranked paths, sensitivity of hyper-parameters, time efficiency of models, etc.

\end{itemize}

\section{Related Work}

\subsection{Social Event Detection}

Early work on social event detection primarily focused on content-based approaches \cite{Wurzer2015Twitter, Wang2017A, Yan2015A}, analyzing only text semantics, while neglecting the social interaction and the heterogeneity of social media data. This oversight hampers the comprehensive capture of information on social media, missing valuable insights from diverse data types \cite{Ren2022Evidential}. Recent developments in social event detection have involved integrating not only textual contents, but also attributes such as user mentions, temporal information, and geolocations. This integration can be achieved through graph models, and then adapted to Graph Neural Networks (GNNs) \cite{Kipf2017Semi, Velickovic2018Graph, Hamilton2017Inductives} to learn node features from the homogeneous graph structure. For example, \cite{Cao2021Knowledge, Ren2022From, peng2022reinforced} built homogeneous text graphs in which they add edges between texts sharing common elements, and then use Graph Attention Networks \cite{Velickovic2018Graph} for neighbor aggregation to gain comprehensive features. Ren et al. \cite{Ren2022Evidential, Ren2023Uncertainty} constructed the text graph from three views: hashtag, entity, and user, building three homogeneous graphs to examine social media data from different perspectives. However, these methods either aggregate different types of node information onto text nodes or construct separate graphs based on different types of nodes, thus neglecting the rich relations between different types of attributes.

\subsection{Meta-Path-Based Methods on HINs}

HINs allow to incorporate diverse types of information flexible, such as text, user interactions, and spatial data, enabling a more comprehensive understanding of events within rich social contexts. Selecting meta-paths from the HINs can be done in three general ways. (1) \emph{Using external knowledge, which is straightforward, but relies heavily on expert experience and human efforts} \cite{Sun2013PathSelClus, Peng2019Fine, wang2019heterogeneous, hao2021streaming, peng2022reinforced}. (2) \emph{Incorporating all meta-paths within a given length.} For instance, Yun et al. \cite{Seongjun2019Graph} introduced Graph Transformer Networks (GTN) to address the limitations of traditional GNNs designed for learning node representations in homogeneous graphs. The GTN model computes an adjacency matrix of meta-paths through matrix multiplications of adjacency matrices, thus eliminating the need for predefined domain knowledge. However, this approach implicitly enumerates all meta-path instances starting from all nodes, which is very time-consuming. In contrast, Li et al. \cite{li2021graphmse} and Qiu et al. \cite{Qiu2024an} used breadth-first search to sample instances of each meta-path, thus reducing the computational overhead by using only a portion of the training data. (3) \emph{Selecting partial meta-paths with automatic mechanisms} \cite{li2023differentiable, ding2021diffmg, ning2022automatic}. For example, Zhong et al. \cite{zhong2022Personalised} designed a reinforcement learning (RL) model that can personalise the meta-paths to each node rather than general ones for each node type. However, RL models are hard to train and have convergence issues. In contrast, the simplicity of parameter-free sampling strategies for important meta-path selection offers an advantage in model convergence.
Li et al. \cite{Li2023long} introduced a two-step training approach that uses progressive sampling to initially select the top-$K$ meta-paths, thus reducing the search space. Nevertheless, selecting an effective meta-path set through HINs for social event detection remains challenging.

\begin{figure*}[h]
\centering
\includegraphics[width=1\textwidth]{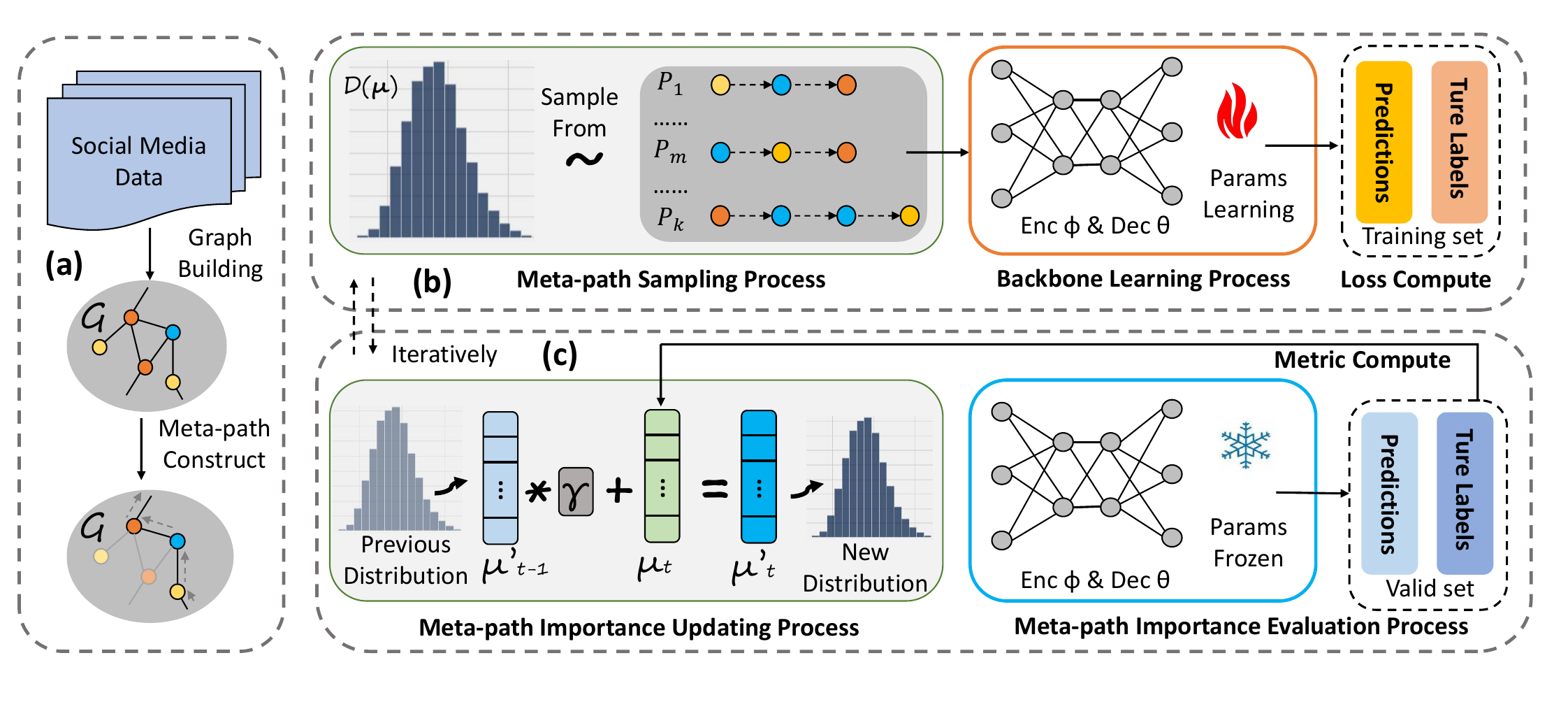}
\caption{The framework of our proposed Meta-path Learning To Sample (LTS) model. Step (a) consists of graph building and meta-path feature construction. Step (b) is model training. Step (c), is meta-path importance evaluation and updating, which are non-parametric. Note that steps (b) and (c) are performed alternatively. }
\label{fig:framework}
\end{figure*}

\section{Preliminaries}

\noindent \textbf{Heterogeneous Information Networks (HINs).} HINs are graph-based data structures that comprise multiple types of nodes and edges. Formally, an HIN can be represented as $\mathcal{G}=(V, E, \mathcal{T}^{v} , \mathcal{T}^{e})$, where $V$ and $E$ denote the sets of nodes and edges, respectively. $\mathcal{T}^{v}$ and $\mathcal{T}^{e}$ represent the sets of nodes and edge types, where $\left | \mathcal{T}^{v} \right | > 1$ and $\left | \mathcal{T}^{e} \right | > 1$. Each node in the network is associated with a unique node type, and similarly, each edge is associated with a unique edge type.  When $\left | \mathcal{T}^{e} \right | = \left | \mathcal{T}^{v} \right | = 1 $, the graph simplifies into a homogeneous structure. HINs offer a richer, more expressive representation of complex relationships between attributes, offering enhanced contextual information and flexibility in modeling.

\vspace{2mm}
\noindent \textbf{Meta-Paths.} A meta-path is a sequence of node types in HINs that represents the specific semantic relationship or path pattern. Formally, a meta-path can be presented as $p = c_{1}\overset{e_{12}}{\rightarrow} c_{2}\overset{e_{23}}{\rightarrow}... \overset{e_{h(h+1)}}{\rightarrow} c_{h+1}$ (abbreviated as $c_{1},c_{2},...,c_{h+1}$), where $h$ is the number of hops. 
Meta-paths enable path-based analysis in HINs, allowing researchers to explore various types of paths connecting different types of nodes. This analysis can uncover latent relationships, patterns, and structures within the network.
For example, in a social network graph, a meta-path ``\emph{Text - User - Text (TUT)}'' represents a path where two message are posted by the same user.

\section{Methodology}

In this section, we present our end-to-end Learning To Sample (LTS) framework, as illustrated in Figure~\ref{fig:framework}. We first construct a heterogeneous information network that includes not only user, text, and entity nodes, but also the connections between them. During the social event detection model training process (b), the top-$K$ most important meta-path features are sampled from the distribution $\mathcal{D} (\boldsymbol{\mu})$. After identifying these key meta-paths, we reduce the exponentially growing number of meta-path combinations to a finite set of highly influential ones. The selected meta-path features and their corresponding weights are then used in training our social event detection model. Path importance weights are updated based on accuracy for classification, in a weighted accumulation manner. Steps (b) and (c) are executed alternately. The overall algorithm is shown in Algorithm \ref{alg}.
The organization of this section is as follows: we begin by outlining the construction of meta-path features (in Section~\ref{sec:meta-path_construction}), and then we provide an overview of the model training process (in Section \ref{sec:SED_model_training}). Afterwards, we offer a detailed explanation of the core learnable sampling components of LTS (in Section \ref{sec: meta-path_searching}), including meta-path sampling strategies (in Section~\ref{sec: meta-path_sampling_strategies}) and methods for evaluating and updating meta-path importance (in Section~\ref{sec: meta-path_importance_evaluation}).

\subsection{Constructing the Meta-Path Features}
\label{sec:meta-path_construction}

A meta-path comprises nodes and edges. In the process of constructing meta-path features, we transform texts, users, and entities into three distinct node categories. Texts serve as the primary carriers of content, while entities represent concepts extracted from the texts. Specifically, we preprocess the texts to remove HTTP links, emojis, and stop-words before being processed for textual feature extraction.

In social media, temporal information captures the time-related aspects of text data, while location offers spatial context, indicating where the events are happening. Therefore, the graph representation is structured as follows: text node features consist of the embedding of the text and of temporal information, while user node features are comprised of filtered word and location embeddings, and entity node features are one-hot vectors representing one or more entities. More formally, the nodes corresponding to text $\mathbf{F}_T$, users $\mathbf{F}_U$ and entities $\mathbf{F}_E$ are defined as follows:

\begin{equation}
\label{eq:node_fts}
\begin{aligned}
    & \mathbf{F}_T = \psi_{cat}(\mathbf{F}_{txt},\mathbf{F}_{temporal}), \\
    & \mathbf{F}_U = \psi_{cat}(\mathbf{F}_{filter\_words},\mathbf{F}_{loc}), \\
    & \mathbf{F}_E = \left\{
        v \in \{0,1\}^n
        \right\},
\end{aligned}
\end{equation}
where $\psi_{cat}(\cdot,\cdots)$ presents the concatenation function. $n$ is the total number of entities.
Moreover, we identify five distinct types of edges between nodes: user--text, text--user, text--entity, entity--text, and user--user. These edges capture the relationships and the interactions between different attributes within the graph.
The model aims to automatically discover the most important meta-paths across the meta-path search space $\mathcal{P}=\{P^1,P^2,\cdots,P^M\}$, where $M$ represents the total number of candidate meta-paths, eliminating the need for human intervention and thereby enhancing the performance through a diverse modeling that comprehensively captures the complex associations within social events. 
To mitigate the increased computational cost resulting from the growing number of meta-path instances, we use the same calculation method for obtaining raw meta-path features as in SeHGNN \cite{yang2023simple}. This involves multiplying the final node features by the normalized adjacency matrix for each meta-path. Formally, the feature of $h$-hop meta-path $m$ can be fetched through neighbor aggregation as follows:

\begin{equation}
\label{eq:aggr}
\mathbf{X}^m = \mathbf{A}^{(c_1, c_2)}\mathbf{A}^{(c_2, c_3)}\cdots \mathbf{A}^{(c_i, c_{i+1})}\cdots \mathbf{A}^{(c_h, c_{h+1})}\mathbf{F}^{(c_{h+1})}.
\end{equation}

\noindent where $\mathbf{A}^{(c_i, c_{i+1})}$ denotes the row-normalized adjacency matrix between node $c_i$ and $c_{i+1}$. $\mathbf{F}^{(c_i)} \in \{\mathbf{F}^{(c_i)}_T,\mathbf{F}^{(c_i)}_U,\mathbf{F}^{(c_i)}_E\}$ is the raw features of node $c_i$, where the value of $\mathbf{F}^{(c_i)}$ depends on its node type.

\begin{algorithm}[!tb]
	\DontPrintSemicolon
	\SetAlgoLined
	
	\KwInput{Social event detection dataset for model training and meta-path importance validation $D=\{D_t,D_v\}$, encoder / decoder structures / initial parameters ($\phi$ and $\theta$), initial meta-path importance vector $\boldsymbol{\mu}$, training iterations N}
	\KwOutput{Encoder and decoder parameters $\phi^{*}$ and $\theta^{*}$, the optimized meta-path importance vector $\boldsymbol{\mu}^{*}$}
        \BlankLine
        \BlankLine
	Build the graph and construct meta-path features. Initialize the model parameters $\phi$ and $\theta$ and the meta-parameter $\boldsymbol{\mu}$ \\
 
	\While{not converged}{

        \textbf{Backbone-training}: \\
	    \For{n=1 to N}{
            Sample partial meta-paths for training: \\
            $p^m \sim \mathcal{D} (\boldsymbol{\mu}^*), \text{if } |\mathcal{P}'| < k$ \\
    	Optimize task objective: \\
       \scalebox{1.0}{
       $\begin{aligned}
        \theta^*, \phi^* & =  \arg \min _{\theta, \phi} \sum_{\mathbf{X}, y \in D_t} \mathcal{L} \left(y, f_{\theta}\left(g_\phi \left(\mathbf{X}\right) \circ \mathcal{P}' \right)\right), \\
        & \hspace{-8mm} \text { s.t. } \mathcal{P}' = \left\{ P^m \in \{0,1\}^M \mid \sum_{m=1}^M P^m = k, \; P^m \sim \mathcal{D}(\boldsymbol{\mu}^*) \right\},
        \end{aligned}$
       }
	    }
        $\theta \leftarrow \theta^*$, $\zeta \leftarrow \zeta^*$ \\

        \textbf{$\boldsymbol{\mu}$-validation \& $\boldsymbol{\mu}$-updating}: \\
        \scalebox{1.0}{$\mu^m_t = \frac{\sum_{\mathbf{X}, y \in D_v} \arg\max(f_{\theta}\left(g_\phi \left(\mathbf{X}\right) \circ \mathcal{P}'_m \right)) \hat{\oplus}y}{\sum_{\mathbf{X}, y \in D_v} 1}, \text{for } \forall \mu \in \left(\mu^1, \ldots, \mu^M\right)$}\\
        $\boldsymbol{\mu}'_t = \boldsymbol{\mu}'_{t-1} * \gamma + \boldsymbol{\mu}_t$ \\
        $\text{Norm}(\boldsymbol{\mu}'_t)$ with Eq.~\eqref{eq:norm_softmax} or Eq.~\eqref{eq:norm_p}\\
	}
	\caption{The Learning To Sample (LTS) Framework.}\label{alg}
\end{algorithm}

\subsection {Model Training}
\label{sec:SED_model_training}

An iterative process is designed to alternately update the parameters of the social event classifier and the meta-path weights. Notably, when updating the meta-path weights, the backbone network parameters are frozen; similarly, during backbone model training, the meta-path importance weights remain unchanged. The optimization problem can be formally presented as follows:

\begin{equation}
\label{eq:meta}
\scalebox{1.0}{$
\begin{aligned}
\theta^*, \phi^* & =  \arg \min _{\theta, \phi} \sum_{\mathbf{\textbf{X}}, y \in D_t} 
\mathcal{L} \left(y, f_{\theta}\left(g_\phi \left(\mathbf{\textbf{X}}\right) \circ \mathcal{P}' \right)\right), \\
& \text { s.t. } \mathcal{P}' = \left\{ P^m \in \{0,1\}^M \mid \sum_{m=1}^M P^m = k, \; P^m \sim \mathcal{D}(\boldsymbol{\mu}^*) \right\},
\end{aligned}
$}
\end{equation}

\noindent where $\mathbf{X} =\{\mathbf{X}^1,\mathbf{X}^2,\cdots,\mathbf{X}^M\}$ presents all sampled meta-path features, $y$ is the ground-truth and $D_t$ denotes the training set data. The function $g_\phi(\cdot)$ parameterized by $\phi$ serves as an encoder that projects meta-path features from their original space into a more semantic feature space.
The vector $\mathcal{P}'$ is a one-hot vector where each value is either 0 or 1. A value of 1 indicates that the corresponding meta-path is selected. The selection of meta-paths for $\mathcal{P}'$ is determined by sampling from the current optimized distribution $\mathcal{D} (\boldsymbol{\mu^*})$, where $\boldsymbol{\mu^*}$ is the current meta-path weight vector.
$k$ represents the number of sampled meta-paths. $\circ$ is the Hadamard product operation. The function $f_\theta$ is a learnable social event classifier parameterized by $\theta$. It can also be understood as a decoder. This process is repeated until the termination condition is reached. Finally, the meta-paths are sorted by their weights, and the top-$K$ meta-paths are selected for evaluation.
Notably, before optimizing the decoder $f_\theta$, we incorporate meta-path importance by applying softmax normalization to weigh each meta-path feature, resulting in $\hat{\mathbf{X}^m}$, the intermediate semantic features fed into the decoder. Formally, we have the following equation:

\begin{equation}
\label{eq:weight_fts}
 \hat{\mathbf{X}^m} = \frac{\exp(\mu^m)}{\sum^M_{j=1}\exp(\mu^j)} g_\phi(\mathbf{X}^m),
\end{equation}

\subsection{Meta-Path Searching}
\label{sec: meta-path_searching}
Longer meta-paths include more correlations and information, but also introduce noise and exponentially increase the number of node combinations, leading to higher computational costs. Therefore, how to find information-rich and important meta-path through limited searches is important. Below, we detail our algorithm for this, which splits the meta-path searching into two parts: (1) sampling strategies, and (2) importance evaluation and updating the meta-paths.

\subsubsection{Meta-Path Sampling Strategies}
\label{sec: meta-path_sampling_strategies}

Meta-paths are different with various degrees of importance. The important meta-path that contributes more to the final task ought to be given larger importance, which in turn should make it rank higher. The goal of meta-path sampling is to find an appropriate set of meta-paths for evaluation and updating. In this section, we explore five different sampling strategies: (1) Random Sampling, (2) Multinomial Distribution Sampling, (3) Decayed $\epsilon$-greedy Sampling, (4) Multinomial $\epsilon$-greedy Sampling, and (5) Multinomial Encouraged $\epsilon$-greedy Sampling.

\vspace{1mm}
\noindent\textbf{Random Sampling.} Random Sampling is a sampling strategy to obtain a vector from the uniform distribution of $\{$0,1$\}$. This sampling strategy assigns each meta-path the same probability to be selected during training.

\vspace{-4mm}
\begin{equation}
\label{eq:uniform_dist}
\mathcal{D} (\boldsymbol{\mu}) = U\{0,1\}
\end{equation}

\vspace{1mm}
\noindent\textbf{Multinomial Distribution Sampling.} As aforementioned, we maintain an $M$-length meta-path weight vector $\boldsymbol{\mu}$. For multinomial distribution sampling, each element $\mu^j$ inside represents the probability of the current meta-path (a.k.a. the importance of each path) to be sampled. The summation of all $\mu^j$ equals to 1.

\begin{equation}
\label{eq:multinomial_dist}
\begin{aligned}
\mathcal{D} (\boldsymbol{\mu}) & : p(x \mid \boldsymbol{\mu})=\prod_{j=1}^M \mu^j_{\mathbb{I}(x=j)}, x \in\{1, \ldots, M\}, \\
& \boldsymbol{\mu}=\left[\mu^1, \mu^2, \ldots, \mu^M\right]^{\mathrm{T}}, \sum_{j=1}^M \mu^j = 1.
\end{aligned}
\end{equation}

\noindent where $\mathbb{I}$ is the indicator function. When $x=j$, $\prod_{j=1}^M \mu^j_{\mathbb{I}(x=j)}$ simplifies to $\mu^j$. Multinomial distribution sampling increases the probability of selecting meta-paths with higher weights, while still allowing for the possibility of choosing such with lower weights. This helps the model potentially identify important meta-paths that may have lower weights initially, but could become important later on.

\vspace{1mm}
\noindent\textbf{Decayed $\epsilon$-greedy Sampling.} 
$\epsilon$-greedy sampling is a simple yet effective sampling strategy, which originated in Reinforcement Learning~\cite{mnih2015human}. In the ordinary $\epsilon$-greedy sampling, $\epsilon$ is a preset constant between 0 and 1. Each time a sampling is required, a random number $\alpha \in (0,1)$ will be generated from a uniform distribution. If $\alpha<\epsilon$, a random choice will be made; otherwise, the current best choice will be returned. The $\epsilon$-greedy sampling is defined as follows:

\begin{equation}
\label{eq:epsilon_greedy}
\mathcal{D} (\boldsymbol{\mu}) =
\begin{cases}
U\{0,1\} , & \text{if } \alpha < \epsilon \\
\mathop{\arg\max}\limits_{\boldsymbol{\mu}}, & \text{if } \alpha \geq \epsilon
\end{cases}
\ ,\ \text{where } \alpha
\sim U(0,1).
\end{equation}

Similarly to multinomial distribution sampling, $\epsilon$-greedy sampling offers higher chance to meta-paths with higher importance. However, in a more explicit and controllable way, it maintains an $\epsilon$ threshold that gives certain chances to other options.
In order to cool-down the random sampling process as the training goes on, we also propose a variant of ordinary $\epsilon$-greedy sampling for meta-path searching, named Decayed $\epsilon$-greedy Sampling, which decays the value of $\epsilon$ as the training iterations increase. It is defined as follows:

\begin{equation}
\label{eq:epsilon_decay}
\epsilon_t = \epsilon_0 \cdot (\beta)^t,
\end{equation}
where $\epsilon_0$ is the initial value of $\epsilon_t$ and $\beta$ is the decay factor at each training iteration $t$. $(a)^b$ represents the power operation. While $\beta=1$, the decayed $\epsilon$-greedy sampling degrades to ordinary $\epsilon$-greedy sampling.
The $\epsilon_t$ can be gradually decreased over time to allow for more exploration in the early stages, while relying more on meta-paths with higher importance in later stages to progressively converge to the optimal solution.

\vspace{1mm}
\noindent\textbf{Multinomial $\epsilon$-greedy Sampling.}  $\epsilon$-greedy Sampling has a deterministic part of always choosing the current best options when $\alpha \geq \epsilon$. In this part, some randomness can be introduced to increase the probability to choose other options, which may help the model converge to a better point. Following the similar vine, we combine the idea of multinomial distribution sampling with decayed $\epsilon$-greedy sampling to form the multinomial $\epsilon$-greedy Sampling. In this sampling, the random selection is kept when the $\alpha$ is smaller than the $\epsilon_t$; however, when $\alpha$ is larger than $\epsilon_t$, making the best choice will be substituted with multinomial distribution sampling. Intuitively, it aims to select `important' meta-paths, but offers more chance to actions with smaller importance, in a softer manner.

\begin{equation}
\label{eq:M_epsilon}
\mathcal{D} (\boldsymbol{\mu}) =
\begin{cases}
U\{0,1\} , & \text{if } \alpha < \epsilon_t \\
p(x \mid \boldsymbol{\mu}), & \text{if } \alpha \geq \epsilon_t
\end{cases}
\ ,\ \text{where } \alpha
\sim U(0,1).
\end{equation}

If the currently high-weight paths are not necessarily globally optimal and the weight of the optimal solution is relatively low, the $\epsilon_t$ setting can help avoid early convergence to local optima. In this case, the random exploration component of multinomial $\epsilon$-greedy sampling helps better discovery of the optimal solution.

\vspace{1mm}
\noindent\textbf{Multinomial Encouraged $\epsilon$-greedy Sampling.} Different from multinomial $\epsilon$-greedy sampling, multinomial encouraged $\epsilon$-greedy sampling further encourages the selection of meta-paths with small importance when $\alpha$ is smaller than $\epsilon_t$.

\begin{equation}
\label{eq:ME_epsilon}
\mathcal{D} (\boldsymbol{\mu}) =
\begin{cases}
p(x \mid \boldsymbol{1-\mu}), & \text{if } \alpha < \epsilon_t \\
p(x \mid \boldsymbol{\mu}), & \text{if } \alpha \geq \epsilon_t
\end{cases}
\ ,\ \text{where } \alpha
\sim U(0,1).
\end{equation}

By assigning higher probability to meta-paths with smaller importance during the exploration phase ($\alpha < \epsilon_t$), this sampling approach helps to uncover potentially important paths that may not be immediately obvious. 

\subsubsection{Meta-Path Importance Evaluation and Update Process}
\label{sec: meta-path_importance_evaluation}

Evaluating and updating the importance of meta-paths accurately is critical. In this subsection, we will detail the processes for initializing, evaluating, updating, and normalizing meta-path importance based on the given metric. This approach directly links the most relevant metric to meta-path importance and is non-parametric, meaning it does not require any learning process.

\vspace{1mm}
\noindent\textbf{Initialization of the Meta-Path Importance.}  The meta-path weight vector is $\boldsymbol{\mu}=\left[\mu^1, \mu^2, \ldots, \mu^M\right]^{\mathrm{T}}$. Each meta-path importance is initialized with an average value, so that the sum of all meta-path importances is 1:
%
\begin{equation}
\label{eq:init_mu}
\sum_{j=1}^M \mu^j = 1, \text{and } \mu^j = \frac{1}{M}.
\end{equation}

\vspace{1mm}
\noindent\textbf{Evaluation of the Meta-Path Importance.} During model training, we use a separate validation process to evaluate the contribution of each meta-path. In this validation process, each meta-path from the sampled set is individually passed through the model to assess its impact on model performance.
We propose to directly use the final metric for evaluating meta-path importance, such as accuracy for classification tasks. As the final metric, it serves as the most direct and intuitive measure of contribution. Specifically, we have:

\begin{equation}
\label{eq:metric}
\mu^m_t = \frac{\sum_{\mathbf{X}, y \in D_v} \arg\max(f_{\theta}\left(g_\phi \left(\mathbf{X}\right) \circ \mathcal{P}'_m \right)) \hat{\oplus}y}{\sum_{\mathbf{X}, y \in D_v} 1},
\end{equation}
where $\mu^m_t$ denotes the contribution value for the path $m$ in the $t$-th training iteration. $\mathcal{P}'_m$ is path $m$ within the sampled path set $\mathcal{P}'$, so we have $\mathcal{P}'_m \in \mathcal{P}'$. Notation $\hat{\oplus}$ represents the inverse of xor operation: return 1 if the predicated class equals to the ground-truth; otherwise, 0 will be returned.

\vspace{1mm}
\noindent\textbf{Updating the Meta-Path Importance.} We design the update of the path importance as a weighted accumulation process: the past values will be gradually decreased as the old ones may not be as accurate as the new ones; so, the newer evaluation should have relatively larger attention. For meta-path weight vector $\boldsymbol{\mu_t}$, we have:
%
\begin{equation}
\label{eq:update}
\boldsymbol{{\mu}_t}' = \boldsymbol{{\mu}_{t-1}}' * \gamma + \boldsymbol{\mu}_t,
\end{equation}
where $\boldsymbol{\mu}_t$ is the immediate values of path contributions and 
$\boldsymbol{\mu}'_t$ is the updated contributions for training iteration $t$. $\gamma$ is the discount factor of previous meta-path importance that is a value between 0 and 1. A smaller value of $\gamma$ will make the algorithm put more emphasis on immediate contributions, while a larger value of $\gamma$ will make the algorithm consider longer-term contributions in the past.

\vspace{1mm}
\noindent\textbf{Normalization of the Meta-Path Importance.} The normalization can guarantee that the path importance values fall in the range of (0,1), which helps the values avoid infinite increments. It can further facilitate probability interpretation and sampling. We propose two variants for the meta-path importance normalization: softmax normalization and $p$-norm normalization. The softmax function~\cite{Krizhevsky2012ImageNet} guarantees that the probability interpretation by summing of all $\mu$ to 1 on the power of $e$.
%
\begin{equation}
\label{eq:norm_softmax}
\mu^{m}_t = \frac{\exp(\mu^{m}_t)}{\sum^M_{j=1}\exp(\mu^j_t)},
\end{equation}

For $p$-norm normalization, we have:

\begin{equation}
\label{eq:norm_p}
\mu^{m}_t=\frac{\mu^{m}_t}{\sum_{j=1}^M \|\mu^j\|_p},
\end{equation}
where $\|\cdot\|_p$ denotes the $p$-norm operation.

\begin{table*}[h]
\setlength\tabcolsep{3pt}
    \centering
    \caption{Overall performance comparison. The best results for each column are in bold.}
    \scalebox{0.88}{
    \begin{tabular}{l|cc|cc|cc|cc|c}
\hline
Datasets   & \multicolumn{2}{c|}{CrisisLexT6}                & \multicolumn{2}{c|}{Kawarith6}                  & \multicolumn{2}{c|}{Twitter2012}                & \multicolumn{2}{c|}{Twitter2018}                            &         \\ \hline
Models     & Micro F1               & Macro F1               & Micro F1               & Macro F1               & Micro F1               & Macro F1               & Micro F1               & \multicolumn{1}{c|}{Macro F1}      & P-Value \\ \hline
TF-IDF     & 0.9365±0.0015          & 0.9356±0.0016          & 0.9287±0.0097          & 0.9249±0.0110          & 0.6789±0.0026          & 0.3405±0.0035          & 0.4259±0.0060          & \multicolumn{1}{c|}{0.2000±0.0089} & 1.59e-2 \\
Word2Vec   & 0.7948±0.0043          & 0.7925±0.0034          & 0.6564±0.0080          & 0.5623±0.0091          & 0.5714±0.0054          & 0.2813±0.0047          & 0.5377±0.0047          & \multicolumn{1}{c|}{0.2224±0.0099} & 1.03e-4 \\
FastText   & 0.9387±0.0009          & 0.9380±0.0009          & 0.8512±0.0019          & 0.8206±0.0032          & 0.1725±0.0000          & 0.0070±0.0000          & 0.0106±0.0001          & \multicolumn{1}{c|}{0.0056±0.0001} & 8.16e-3 \\
FinEvent   & -                      & -                      & 0.9259±0.0087          & 0.9136±0.0132          & -                      & -                      & -                      & -                                 & -       \\
BERT       & 0.8495±0.0038          & 0.8468±0.0036          & 0.7695±0.0092          & 0.7502±0.0101          & 0.6889±0.0032          & 0.5158±0.0052          & 0.5545±0.0038          & \multicolumn{1}{c|}{0.3200±0.0091} & 1.94e-4 \\
GraphMSE   & 0.9100±0.0032          & 0.9094±0.0031          & 0.9470±0.0008          & 0.9400±0.0003          & 0.8057±0.0046          & 0.6716±0.0094          & 0.7671±0.0034          & \multicolumn{1}{c|}{0.6621±0.0071} & 1.87e-3 \\
ETGNN      & -                      & -                      & -                      & -                      & 0.8480±0.0000          & 0.7565±0.0000          & -                      & -                                 & -       \\
HGT        & 0.9192±0.0034          & 0.9178±0.0035          & 0.9193±0.0036          & 0.9118±0.0038          & 0.7111±0.0036          & 0.5841±0.0036          & \textbf{0.8046±0.0026} & 0.6892±0.0095                     & 1.49e-2 \\
KPGNN      & 0.4451±0.0339          & 0.4143±0.7392          & 0.7863±0.0171          & 0.7691±0.0106          & -                      & -                      & -                      & -                                 & 2.15e-2 \\
GraphHAM   & 0.9218±0.0027          & 0.9213±0.0023          & 0.9510±0.0041          & \textbf{0.9457±0.0053} & 0.8414±0.0013          & 0.7100±0.0040          & -                      & -                                 & 1.84e-2 \\ \hline
LTS (Ours) & \textbf{0.9586±0.0009} & \textbf{0.9576±0.0009} & \textbf{0.9523±0.0020} & 0.9455±0.0023          & \textbf{0.8632±0.0009} & \textbf{0.7652±0.0030} & 0.8021±0.0030          & \textbf{0.7179±0.0022}            & -       \\ \hline
\end{tabular}}
    \label{tab:overall}
\end{table*}

\section{Experiments and Evaluation}

In this section, we first introduce the datasets, the evaluation metrics, and the graph construction methods. Details about the baseline models and the implementation specifics are provided in the Appendix. We then conduct experiments to address the following research questions:

\begin{itemize}
\item \textbf{Q1:} Is the proposed LTS model comparable in performance to other strong baseline models?

\item \textbf{Q2:} What are the most effective meta-path sampling strategies?

\item \textbf{Q3:} How to normalize the meta-path importance weights?

\item \textbf{Q4:} How do the top-ranked paths impact the performance of the social event detector?

\item \textbf{Q5:} What is the impact of the values of hyper-parameters such as hidden neuron size, discount factor of previous meta-path importance, and number of hops?

\item \textbf{Q6:} What is the time efficiency of the proposed LTS model?

\end{itemize}

\subsection{Datasets, Evaluation Metrics, and Graph Building Structures}

We conduct experiments on four datasets: CrisisLexT6 \cite{olteanu2014crisislex}, Kawarith6 \cite{alharbi2021kawarith}, Twitter2012 \cite{James2013Building}, and Twitter2018\cite{Beatrice2020a}. CrisisLexT6 contains 18,157 messages belonging to six unique event classes; Kawarith6 contains 4,860 messages belonging to six unique event classes; Twitter2012 contains 68,841 messages from 503 unique event classes; Twitter2018 contains 64,516 messages from 257 unique event classes.

For evaluation, we use Micro-F1 and Macro-F1. Micro-F1 evaluates the overall performance in classifying instances across all classes, while Macro-F1 checks each class individually before averaging these scores.

The LTS method can be applied to other tasks if HINs are able to be constructed and the meta-path features can be obtained. In our work, we focus on harnessing the potential of social media data by transforming it into a structured graph using HINs. We introduce three node types: Text Contents (T), Users (U), and Entities (E). Additionally, we define five edge types to capture various connections: User--Text, Text--User, Text--Entity, Entity--Text, and User--User.

\subsection{Overall Performance (Q1)}
The compared models include TF-IDF \cite{akiko2003an}, Word2Vec \cite{mikolov2013efficient}, FastText \cite{armand2017bag}\footnote{The Twitter2012 and Twitter2018 datasets contain a large number of event classes and exhibit data imbalance, with some tweets appearing only once. While FastText excels at learning from frequently occurring words due to its bag-of-words based model, this is actually detrimental when applied to the Twitter2012 and Twitter2018 datasets.}, FinEvent \cite{peng2022reinforced}\footnote{We  also tried FinEvent on CrisisLexT6, Twitter2012 and Twitter2018, but it adopts GAT for message passing that is very memory-intensive and casted out-of-memory errors on these three datasets.}, BERT \cite{kenton2019bert}, GraphMSE \cite{li2021graphmse}, ETGNN \cite{Ren2022Evidential}\footnote{We report the performance of ETGNN from the paper~\cite{Ren2022Evidential}, they evaluated the model on CrisisLexT26 and Kawarith7, which have different data scales with ours.}, KPGNN \cite{Cao2021Knowledge}\footnote{KPGNN encounter OOM errors on the Twitter2012 and Twitter2018, and thus their results are not included.}, HGT \cite{Hu2020Heterogeneous} and GraphHAM \cite{Qiu2024an}.
Table~\ref{tab:overall} shows the performance of the compared models and of our proposed LTS model across the four datasets. The LTS model consistently outperforms the baselines and achieves state-of-the-art performance in most of the cases. The p-value analysis demonstrates that the performance improvements achieved by our proposed LTS method over the baseline models are statistically significant (p < 0.05) across various datasets.
On the Twitter2012 datasset, notably ETGNN is the second best model and it has much stronger performance than the other models. Our proposed LTS model outperforms ETGNN by 1.79\% on Micro-F1 and 1.15\% on Macro-F1. On CrisisLexT6, LTS performs particularly better than the others, e.g., compared to the FastText model (the second best), with 2.12\% improvement for Micro-F1 and 2.09\% for Macro-F1. In the Kawarith6 dataset, the Macro F1 of the proposed LTS is 0.0002 lower than that of GraphHAM. This is because the GraphHAM model uses hyperbolic space representation to reduce the distortion caused by embedding tree-structured data in the Euclidean space, thereby improving model performance. If GraphHAM were to use Euclidean space, as LTS does, its performance would drop substantially, resulting in a performance lower than that of LTS.

\subsection{Sampling Strategies (Q2)}

In this experiment, we evaluate the impact of different sampling strategies on our LTS model. The sampling strategies include (1) Random Sampling (RS), (2) Multinomial Distribution Sampling (MDS), (3) Ordinary $\epsilon$-greedy Sampling (O-$\epsilon$S), which is a degraded version of D-$\epsilon$S,  (4) Decayed $\epsilon$-greedy Sampling (D-$\epsilon$S), (5) Multinomial $\epsilon$-greedy Sampling (M-$\epsilon$S), and (6) Multinomial Encouraged $\epsilon$-greedy Sampling (ME-$\epsilon$S). The results are shown in Table~\ref{tab:sampling}. Across the three datasets, ME-$\epsilon$S achieves the best results on both measures (two datasets). On Twitter2012, the D-$\epsilon$S achieves the best results, but the ME-$\epsilon$S also achieves a comparable good results (second best in terms of Micro F1). We believe 
that this is because the ME-$\epsilon$S increases the extra randomness inside the D-$\epsilon$S, changing the deterministic choice of current best options into an action with some probabilities to choose some current sub-optimal solution. This choice may eventually result in better performance. Nevertheless, for ME-$\epsilon$S, MDS have been injected into it either when $\alpha < \epsilon$ or $\alpha \geq \epsilon$. It may form a relatively strong constraint, which incurs sub-optimal sampling solutions.

\begin{table}[h]
    \centering
    \caption{The evaluation of different sampling strategies. The best results in each column are in bold.}
    \scalebox{0.78}{
    \begin{tabular}{l|cc|cc|cc}
\hline \hline
Datasets           & \multicolumn{2}{c|}{Twitter2012}                              & \multicolumn{2}{c|}{CrisisLexT6}                              & \multicolumn{2}{c}{Kawarith6}                                                         \\ \hline
Models         & \multicolumn{1}{c|}{Micro F1} & \multicolumn{1}{c|}{Macro F1} & \multicolumn{1}{c|}{Micro F1} & \multicolumn{1}{c|}{Macro F1} & \multicolumn{1}{c|}{Micro F1} & \multicolumn{1}{c}{Macro F1} \\ \hline
RS             & \multicolumn{1}{c|}{0.8671}   & 0.7607                        & \multicolumn{1}{c|}{0.9511}   & 0.9499                        & \multicolumn{1}{c|}{0.9253}   & 0.9157                                                \\
MDS            & \multicolumn{1}{c|}{0.8720}    & 0.7583                        & \multicolumn{1}{c|}{0.9539}   & 0.9528                        & \multicolumn{1}{c|}{0.9304}   & 0.9208                                                \\
O-$\epsilon$S  & \multicolumn{1}{c|}{0.8747}   & 0.7606                        & \multicolumn{1}{c|}{0.9483}   & 0.9471                        & \multicolumn{1}{c|}{0.9253}   & 0.9165                                                \\
D-$\epsilon$S  & \multicolumn{1}{c|}{\textbf{0.8767}}   & \textbf{0.7634}                        & \multicolumn{1}{c|}{0.9518}   & 0.9508                        & \multicolumn{1}{c|}{0.9278}   & 0.9193                                                \\
M-$\epsilon$S  & \multicolumn{1}{c|}{0.8754}   & 0.7578                        & \multicolumn{1}{c|}{\textbf{0.9573}}   & \textbf{0.9568}                        & \multicolumn{1}{c|}{\textbf{0.9407}}   & \textbf{0.9341}                                                \\
ME-$\epsilon$S & \multicolumn{1}{c|}{0.8749}   & 0.7599                        & \multicolumn{1}{c|}{0.9518}   & 0.9505                        & \multicolumn{1}{c|}{0.9330}    & 0.9241                                                \\ \hline \hline
\end{tabular}}
    \label{tab:sampling}
\end{table}

\begin{table}[h]
    \centering
    \caption{Evaluation of different normalization functions for meta-path contribution. The best results per column are in bold.}
    \scalebox{0.78}{
    \begin{tabular}{l|cc|cc|cc}
\hline \hline
Datasets           & \multicolumn{2}{c|}{Twitter2012}                              & \multicolumn{2}{c|}{CrisisLexT6}                              & \multicolumn{2}{c}{Kawarith6}                                                         \\ \hline
Norms         & \multicolumn{1}{c|}{Micro F1} & \multicolumn{1}{c|}{Macro F1} & \multicolumn{1}{c|}{Micro F1} & \multicolumn{1}{c|}{Macro F1} & \multicolumn{1}{c|}{Micro F1} & \multicolumn{1}{c}{Macro F1} \\ \hline
Softmax & \multicolumn{1}{c|}{0.8676}   & 0.7536   & \multicolumn{1}{c|}{0.9525}   & 0.9513   & \multicolumn{1}{c|}{0.9278}   & 0.9203                           \\
L1      & \multicolumn{1}{c|}{\textbf{0.8754}}   & 0.7578   & \multicolumn{1}{c|}{\textbf{0.9573}}   & \textbf{0.9568}   & \multicolumn{1}{c|}{\textbf{0.9407}}   & \textbf{0.9341}                           \\
L2      & \multicolumn{1}{c|}{0.8733}   & \textbf{0.7600}     & \multicolumn{1}{c|}{0.9525}   & 0.9514   & \multicolumn{1}{c|}{0.9330}    & 0.9243                           \\ \hline \hline
\end{tabular}}
    \label{tab:norm}
\end{table}

\subsection{Normalization Function (Q3)}

We also examine the effectiveness of different normalization functions for meta-path importance values, as shown in Table~\ref{tab:norm}.
In general and performing consistently, the L1-normalization of meta-path contribution yields almost all best results across the three datasets and the two evaluation metrics. 
Interestingly, the L2-normalization methods yield slightly worse, but comparable results, while softmax normalization achieves the worst out of these three functions under our settings. We hypothesize that L1-normalization is the most direct normalization to sum up all available values to 1, while L2-normalization and softmax normalization should have the power of 2 and with the base of $e$, which are not as direct as L1-normalization.

\begin{table*}[h]
    \centering
    \caption{The top-10 ranked meta-paths on three datasets.}
    \scalebox{0.78}{
    \begin{tabular}{l|c}
\hline \hline
Datasets    & Top-10 Ranked Meta-paths for each dataset                                                    \\ \hline
Twitter2012 & `TUTETE',`TETE',`TET',`TETUTU',`TE', 'TUTE',`TETUT',`TETET',`T',`TUTETU'   \\
CrisisLexT6 & `TUTETE',`TUTUTU',`TUTUT',`TE',`TUT',`TETUT',`TET',`TETETU',`TETUTU',`TU' \\
Kawarith6   & `TETETE',`TETE',`T',`TETUTU',`TUTU',`TU',`TUTUTU',`TETUT',`TUTET',`TUTUT' \\ \hline \hline
\end{tabular}}
    \label{tab:path_qualitative}
\end{table*}

\subsection{Top-Ranked Important Meta-Paths (Q4)}
\subsubsection{Qualitative Analysis}

We also perform a qualitative analysis of the top-ranked meta-paths by listing them in Table~\ref{tab:path_qualitative}. We can see that \emph{TUTETE} is ranked twice as top-1 on three datasets, e.g., on Twitter2012 and CrisisLexT6, and that the entity nodes are important on all three datasets as the three top-1 paths all end with \emph{E}. It indicates that the entity information is critical for the class prediction task. Also, interestingly, longer paths are preferred. This suggests that longer paths contain more meaningful information for model training.

\subsubsection{Quantitative Analysis }

\begin{table}[h]
    \centering
    \caption{Model performance when gradually deleting the top-ranked levels. ``Del lv'' denotes deleting level and ``Med'' represents medium level.}
    \scalebox{0.78}{
    \begin{tabular}{l|cc|cc|cc}
\hline \hline
Datasets           & \multicolumn{2}{c|}{Twitter2012}                              & \multicolumn{2}{c|}{CrisisLexT6}                              & \multicolumn{2}{c}{Kawarith6}                                                         \\ \hline
Del lv         & \multicolumn{1}{c|}{Micro F1} & \multicolumn{1}{c|}{Macro F1} & \multicolumn{1}{c|}{Micro F1} & \multicolumn{1}{c|}{Macro F1} & \multicolumn{1}{c|}{Micro F1} & \multicolumn{1}{c}{Macro F1} \\ \hline
LTS           & \multicolumn{1}{c|}{0.8754}   & 0.7578   & \multicolumn{1}{c|}{0.9573}   & 0.9568   & \multicolumn{1}{c|}{0.9407}   & 0.9341                           \\
Mild          & \multicolumn{1}{c|}{0.8750}   & 0.7553   & \multicolumn{1}{c|}{0.9558}   & 0.9547   & \multicolumn{1}{c|}{0.9330}   & 0.9234                           \\
Med        & \multicolumn{1}{c|}{0.8676}   & 0.7533   & \multicolumn{1}{c|}{0.9456}   & 0.9447   & \multicolumn{1}{c|}{0.9201}   & 0.9130                           \\
Strong        & \multicolumn{1}{c|}{0.8547}   & 0.7360   & \multicolumn{1}{c|}{0.9146}   & 0.9131   & \multicolumn{1}{c|}{0.9175}   & 0.9075                           \\ \hline \hline
\end{tabular}}
    \label{tab:path_del}
\end{table}

In order to explore the influence of meta-paths on social event detection performance and to demonstrate the effectiveness of LTS that can pick the most important meta-paths, we adopt a leave-out principle to check how the performance changes if we remove meta-paths from the LTS-found meta-path set. We set the meta-path exclusion intense level as mild, medium, and strong to remove two, four and six paths, respectively.
By gradually removing the top-ranked meta-paths in Table~\ref{tab:path_del}, we observe a decrease in model performance as the intensity of removal increases across different datasets and metrics.

\subsection{Sensitivity to Hyper-Parameters (Q5)}

\subsubsection{Sensitivity to Hidden Neuron Size}

We also perform experiments aimed at providing both qualitative and quantitative sensitivity analysis for various hidden neuron sizes within the backbone network.
The upward-and-downward trend depicted in Figure~\ref{fig:sen_hidden} underscores the influence of hidden neuron size. Based on the graph, a hidden neuron size of 512 is recommended, as it consistently yields the best performance across both evaluation measures.

\subsubsection{The Sensitivity of the Discount Factor of Previous Meta-Path Importance $\gamma$}

\begin{figure}[!tb]
\centering
\subfigure{
\begin{minipage}[h]{0.45\linewidth}
\centering
\includegraphics[width=1\textwidth]{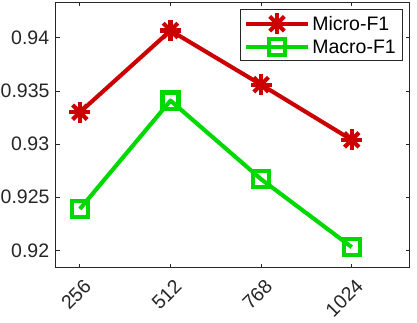}
(a)
\label{fig:sen_hidden}
\end{minipage}%
    }%
\subfigure{
\begin{minipage}[h]{0.45\linewidth}
\centering
\includegraphics[width=1\textwidth]{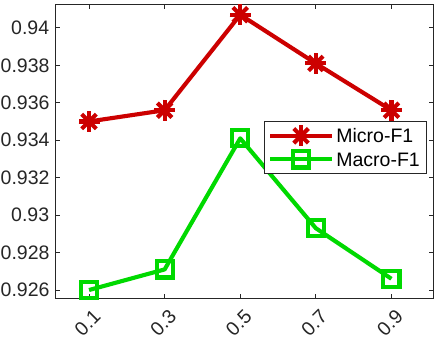}
(b)
\label{fig:sen_gamma}
\end{minipage}%
}%
\centering
\caption{Sensitivity test for (a) hidden neuron size and (b) discount factor of previous meta-path importance $\gamma$.}
\end{figure}

Similarly to the sensitivity to hidden neuron size, we 
examine the changes in model performance w.r.t. the change of discount factor of previous meta-path importance $\gamma$ and draw the figure, as shown in Figure~\ref{fig:sen_gamma}.
The Micro-F1 and Macro-F1 value are also following the trend of going up and down afterwards. When $\gamma=0.5$, it reaches the peak. This may be because retaining too little or too much information from past meta-path importance would not significantly contribute to the selection of meta-paths.

\subsubsection{Sensitivity to Hops}

\begin{table}[h]
    \centering
    \caption{Sensitivity test for the hops on CrisisLexT6 dataset. ``Ave. Speed'' is the average processing speed the unit is processed iterations per second (the higher the better).}
    \scalebox{0.78}{
    \begin{tabular}{l|c|c|c}
\hline \hline
\multicolumn{1}{l|}{Hops } & \multicolumn{1}{l|}{Micro F1} & \multicolumn{1}{l|}{Macro F1} & Ave. Speed \\ \hline
Hops=4                          & 0.9545                        & 0.9532                        & 2.76 iterations/s \\
Hops=5                          & 0.9573                        & 0.9568                        & 3.00 iterations/s \\
Hops=6                          & 0.9580                        & 0.9570                        & 3.10 iterations/s \\
Hops=7                          & 0.9570                        & 0.9559                        & 2.68 iterations/s \\ \hline \hline
\end{tabular}}
    \label{tab:sen_hops}
\end{table}

We also conduct a sensitivity test w.r.t. the number of hops. The more hops, the longer path features have been calculated, and thus inevitably more information will be considered, which grows exponentially as the combination explodes. However, with our proposed learnable sampling, we can constrain the processed path features into a finite set with a constant number of members per training iteration.

As shown in Table~\ref{tab:sen_hops}, we can observe on the dataset, that as the number of hops increases (from 4 to 6), both Micro and Macro F1 scores follow the trend to go up, meaning more useful information has been considered. However, with 7 hops, the increasing trend stops as possibly more noise is introduced, which confuses the model. The average processing speed follows a very stable trend. This suggests that our proposed Learning To Sampling (LTS) framework can effectively reduce the processing time when the combination of meta-path grows excessively.

\subsection{Time Efficiency (Q6)}
We compared the inference time to process test data on three datasets between baselines and LTS. Table~\ref{tab:time} shows an inference time comparison for various baseline models\footnote{The paper on ETGNN \cite{Ren2022Evidential} did not release the code, and thus we are unable to calculate their inference time.}. We can see that our proposed LTS model demonstrates notably low time complexity, outperforming most other models. For example, on the Twitter2012 dataset, LTS only takes 1.60 seconds, which is  faster than the second fast model GraphMSE which takes 2.57s. These results illustrate that LTS not only offers low time complexity, but also maintains robust performance, which makes it an efficient and effective choice across a number of datasets.

\begin{table}[h]
    \centering
    \caption{Inference time comparison between models.}
    \scalebox{0.78}{
    \begin{tabular}{l|c|c|c}
\hline \hline
Models/Datasets & Twitter2012 & CrisisLexT6 & Kawarith6 \\ \hline
TF-IDF          & 11.49s      & 0.21s       & 0.07s     \\
Word2Vec        & 27.34s      & 5.32s       & 1.41s     \\
FastText        & 32.54s      & 2.02s       & 0.14s     \\
FinEvent        & -           & -           & 3.02s     \\
BERT            & 1303.10s    & 312.61s     & 78.42s    \\
GraphMSE        & 2.57s       & 1.01s       & 0.21s     \\
KPGNN           &  -         & 392.06s          & 42.60s              \\
GraphHAM        & 4.07s       & 2.10s       & 0.80s     \\
LTS (Ours)      & 1.60s       & 0.38s       & 0.16s     \\ \hline
\hline \hline
\end{tabular}}
    \label{tab:time}
\end{table}

\section{Conclusion and Future Work}
We studied social event detection in the scope of heterogeneous information networks (HINs) and proposed a simple, yet effective, end-to-end Learning To Sample (LTS) framework for the task. We compared five sampling methods and a non-parametric weighted accumulation evaluation process to effectively identify important meta-paths.
Our extensive experiments and analysis suggest that our proposed LTS model can highlight the important meta-path set to help the backbone model training and inference. LTS also reduces the meta-path searching time via sampling, which demonstrates the efficiency of our proposed framework. 

One potential future direction for further developing our model is to improve the metrics used for updating path importance. Currently, we use the most intuitive and direct metrics for a classification task. However, we plan to consider additional factors, such as loss.
Furthermore, we aim to apply our meta-path searching strategies to other tasks involving HINs and to test the generalization of LTS in different domains. Another important direction is to extend LTS from its current focus of offline data to dynamic data. This would enhance its applicability and robustness in various real-world scenarios.

\bibliographystyle{ACM-Reference-Format}
\bibliography{sample-base}

\appendix
\section{Baseline models}
We compare the proposed LTS model with several cutting-edge methods, including: 

\begin{itemize}

\item Term Frequency-Inverse Document Frequency (TF-IDF) \cite{akiko2003an} is a statistical measure used to evaluate the importance of a word in a document relative to a collection of documents. It is based on how frequently the words appears in the document and inversely proportional to how often the words appear across all textual data.

\item Word2Vec \cite{mikolov2013efficient} is a neural network-based technique used to learn distributed representations of words in a continuous vector space from large text corpora.

\item FastText \cite{armand2017bag} is library for efficient learning of word embeddings and text classification, developed by Facebook AI Research, known for its speed and ability to handle out-of-vocabulary words by utilizing character n-grams.

\item FinEvent \cite{peng2022reinforced} is a powerful model that is specifically designed for social event detection, introducing a multi-agent reinforced weighted multi-relational graph neural network framework to learn discriminative message embeddings.
    
\item Bidirectional Encoder Representations from Transformers (BERT) \cite{kenton2019bert} is a natural language processing (NLP) model that uses a bidirectional Transformer architecture to pre-train deep contextualized word representations on large text corpora, enabling fine-tuning for various downstream tasks.

\item GraphMSE \cite{li2021graphmse} is a meta-path-based method that employs heterogeneous Graph Convolutional Networks (GCNs), utilizing a breadth-first search technique to sample instances of each meta-path, thereby mitigating computational costs.

\item ETGNN \cite{Ren2022Evidential} is another strong model that is specifically designed for social event detection. This model considers temporal awareness and utilizes Dempster-Shafer theory for evidence fusion.

\item KPGNN \cite{Cao2021Knowledge} models complex social information as a unified social graph to better utilize data and explore the expressive power of GNNs for knowledge extraction.

\item HGT \cite{Hu2020Heterogeneous} is a GNN based model designed for modeling large-scale heterogeneous graphs, which can automatically learn the heterogeneous attention between different types of nodes and edges without manually designed meta-paths.

\item GraphHAM \cite{Qiu2024an} investigates social event detection and proposes an efficient heterogeneous information graph learning framework. This framework involves the selection of meta-paths and the use of hyperbolic space to learn information from social media.
\end{itemize}

\section{Implementation details}
In the meta-path feature construction, the text node features comprise 100-dimensional embeddings derived from the text sentences, which are further combined with 2-dimensional temporal information embeddings. On the other hand, user node features are constructed by combining 100-dimensional embeddings of filtered words with 2-dimensional location embeddings. We allocate 70\% of the data as the training set, 10\% as the validation set, and the remaining 20\% as the testing set. Our approach employs a preprocessing step to de-identify the username into numerical ids for all datasets.
We used TF-IDF, word2vec, and BERT to obtain features and then applied a logistic regression for event detection. We report the overall performance of the model on the testing set and report additional analysis experiments on the validation set. The non-parametric meta-path evaluation process is performed on the validation set. Our model takes into account 5 hops for nodes. During the training process, we employ the Adam optimizer with a consistent learning rate of 0.001 across 300 epochs, using a batch size of 10,000. Additionally, we adopt the same Adam optimizer with a fixed learning rate of 0.001 for 300 epochs of training, maintaining a batch size of 10,000. By default, the top influential meta-paths selection is set to k=10 and $\gamma$=0.5. Our encoder architecture comprises one fully connected layers while the decoder has two fully connected layers, incorporating BatchNorm, PReLU non-linearities, and dropout with a rate of 0.5 between them. Residual connections are also incorporated. We set the hidden size of the encoder outputs to 512. By default, the multinomial $\epsilon$-greedy Sampling and L1-norm are adopted. Decayed $\epsilon$ is activated and we set $\epsilon_0$=0.5, $\beta$=0.99. The random seed is 666.

\section{Theoretical Proof}

\textit{Theorem 1}: Let \(\text{Acc}^{j}\) be the accuracy of the \(j\text{-th}\) meta-path, $\mu ^{j}$ be the corresponding meta-path weight, and $\sum_{j=1}^{M} \mu^{j} = 1$, then the expectation of multinomial distribution sampling (MDS)is greater than or equal to the expectation of random sampling (Random):

\[
E_{\text{MDS}} \ge E_{\text{Random}}.
\]

\noindent \textit{Proof 1}: The expectation of random sampling can be expressed as:

\[
E_{\text{Random}} = \sum_{j=1}^{M} \frac{1}{M} \text{Acc}^{j},
\]

\noindent whereas the expectation of multinomial distribution sampling is:

\[
E_{\text{MDS}} = \sum_{j=1}^{M} \mu^{j} \text{Acc}^{j}.
\]

According to the sampling mechanism inherent in multinomial distribution sampling, there exists a positive correlation between the accuracy \(\text{Acc}^{j}\) and the corresponding meta-path weights \(\mu^{j}\). Specifically, this correlation implies that higher accuracy values are associated with higher meta-path weights. Given that the meta-path accuracies and weights are sorted in non-decreasing order:

\[
\text{Acc}^{1} \le \text{Acc}^{2} \le \ldots \le \text{Acc}^{M},
\]

\[
\mu^{1} \le \mu^{2} \le \ldots \le \mu^{M},
\]

Beacuse $\forall$ $\mu ^{j}$, we have $\mu ^{j}> 0$ and $\sum_{j=1}^{M} \mu ^{j} =1$, according to Chebyshew inequality, we have:

\begin{align*}
E_{MDS} = & \sum_{j=1}^{M} \mu ^{j}  Acc^{j} 
\ge \left (\sum_{j=1}^{M} \mu ^{j}\right ) \left ( \sum_{j=1}^{m} \frac{Acc^{j}}{M}  \right ) \\\\
& = \frac{1}{M}\sum_{j=1}^{M }Acc^{j}  = E_{Random}
\end{align*}

Under the assumption that the data are independently and identically distributed, the model's accuracy for identical meta-paths remains largely consistent across the validation and testing sets. Consequently, under this premise, multinomial distribution sampling performs better than Random Sampling.

\vspace{2mm}
\noindent \textit{Theorem 2}: In multinomial $\epsilon$-greedy sampling, as the number of samples $t \to \infty$, the sample proportion $\frac{N_j(t)}{t}$ will converge to its true probability $p_j$.

\vspace{2mm}
\noindent \textit{Proof 2}: In each sampling iteration, $\alpha$ is a random variable drawn from a uniform distribution $U(0,1)$. We divide the sampling process into two phases: 

\noindent (1) exploration phase: when $\alpha < \epsilon_t$, each path is uniformly selected with probability $\frac{1}{M}$.

\noindent (2) exploitation phase: when $\alpha \geq \epsilon_t$, each path $j$ is selected with probability $\mu ^{j}$.

Therefore, in the $t$-th sampling iteration, the total expected number of times path $j$ is chosen $E(N_j(t))$ can be expressed as:

\begin{align*}
E(N_j(t)) &= \epsilon_t \cdot t \cdot \frac{1}{M} + (1 - \epsilon_t) \cdot t \cdot \mu^{j} 
 \\
&= \epsilon_0 \cdot (\beta)^t \cdot t \cdot \frac{1}{M} + (1 - \epsilon_0 \cdot (\beta)^t) \cdot t \cdot \mu^{j} 
\\
&= \epsilon_0 \cdot (\beta)^t \cdot t \cdot \frac{1}{M} + t \cdot \mu^{j} - \epsilon_0 \cdot (\beta)^t \cdot t \cdot \mu^{j}
\\
&= t \cdot \mu^{j} +  \epsilon_0 \cdot (\beta)^t \cdot t \cdot(\frac{1}{M}-\mu^{j})
\end{align*}

Because $\epsilon_0$ and $\beta$ are constant and $\epsilon_0 < 1$, $\beta <1 $. As $t \to \infty$, the term $\epsilon_0 * (\beta)^t \cdot t \cdot(\frac{1}{M}-\mu^{j})\to \infty$. Thus:

\[
E(N_j(t)) \approx t \cdot \mu^{j}
\]

According to the Law of Large Numbers, as $t$ becomes sufficiently large, the sample average will converge to its expected value. Therefore, for each $j$:

\[
\frac{N_j(t)}{t} \to \mu^{j} \quad \text{almost surely}
\]

This means that the proportion of times each meta-path is chosen, as the number of samples increases, will almost surely converge to its true probability $p_j$.

\end{document}